\documentclass[12pt,a4paper,final]{iopart}

\usepackage[utf8]{inputenc}
\usepackage{graphicx}
\usepackage{iopams}
\usepackage{dcolumn}
\usepackage{bm}
\usepackage{amssymb}
\usepackage{setspace}
\usepackage[usenames,dvipsnames,svgnames]{xcolor}
\usepackage[breaklinks=true,colorlinks=true,linkcolor=blue,urlcolor=blue,citecolor=blue]{hyperref}

\newcommand{\ket}[1]{\ensuremath{|\,#1\,\rangle}}

\newcommand{\elem}[3]{\ensuremath{\langle\,#1\,|\,#2\,|\,#3\,\rangle}}

\newcommand{\SI}[2]{\ensuremath{#1 \: \mathrm{#2}}}
\newcommand{\cm}[1]{\SI{#1}{cm^{-1}}}

\newcommand{\wwX}{X\ensuremath{^1\Sigma^+}}
\newcommand{\wwa}{a\ensuremath{^3\Sigma^+}}
\newcommand{\wwb}{b\ensuremath{^3\Pi}}
\newcommand{\wwA}{A\ensuremath{^1\Sigma^+}}
\newcommand{\wwc}{c\ensuremath{^3\Sigma^+}}
\newcommand{\wwB}{B\ensuremath{^1\Pi}}

\newcommand{\omO}{\ensuremath{\Omega=0^+}}
\newcommand{\omI}{\ensuremath{\Omega=1}}

\begin{document}

\title{Prospects for the formation of ultracold polar ground state KCs molecules via an optical process}

\author{D. Borsalino$^{1}$}
\address{$^1$Laboratoire Aim$\acute{e}$ Cotton, CNRS/Universit$\acute{e}$ Paris-Sud/ENS Cachan, Orsay Cedex, France}

\author{R. Vexiau$^{1}$}
\address{$^1$Laboratoire Aim$\acute{e}$ Cotton, CNRS/Universit$\acute{e}$ Paris-Sud/ENS Cachan, Orsay Cedex, France}

\author{M. Aymar$^{1}$}
\address{$^1$Laboratoire Aim$\acute{e}$ Cotton, CNRS/Universit$\acute{e}$ Paris-Sud/ENS Cachan, Orsay Cedex, France}

\author[cor1]{E. Luc-Koenig$^{1}$}
\address{$^1$Laboratoire Aim$\acute{e}$ Cotton, CNRS/Universit$\acute{e}$ Paris-Sud/ENS Cachan, Orsay Cedex, France}

\author[cor1]{O. Dulieu$^{1}$}
\address{$^1$Laboratoire Aim$\acute{e}$ Cotton, CNRS/Universit$\acute{e}$ Paris-Sud/ENS Cachan, Orsay Cedex, France}
\eads{\mailto{olivier.dulieu@u-psud.fr}}

\author{N. Bouloufa-Maafa$^{1,3}$}
\address{$^1$Laboratoire Aim$\acute{e}$ Cotton, CNRS/Universit$\acute{e}$ Paris-Sud/ENS Cachan, Orsay Cedex, France}
\address{$^{3}$ UFR de Physique, Universit\'{e} de Cergy-Pontoise, France}


\date{\today}

\begin{abstract}
Heteronuclear alkali-metal dimers represent the class of molecules of choice for creating samples of ultracold molecules exhibiting an intrinsic large permanent electric dipole moment. Among them, the KCs molecule, with a permanent dipole moment of 1.92~Debye still remains to be observed in ultracold conditions. Based on spectroscopic studies available in the literature completed by accurate quantum chemistry calculations, we propose several optical coherent schemes to create ultracold bosonic and fermionic KCs molecules in their absolute rovibrational ground level, starting from a weakly bound level of their electronic ground state manifold. The processes rely on the existence of convenient electronically excited states allowing an efficient stimulated Raman adiabatic transfer of the level population.
\end{abstract}


\maketitle

\section{Introduction}
\label{sec:intro}

Dilute atomic and molecular gases at ultracold temperatures ($T=E/k_B \ll 1$~millikelvin) offer the fascinating opportunity of long observation time allowing for measurements with unprecedented accuracy. For instance, due to their extremely low relative velocity in such gases, particles have their maximal presence probability at large mutual distances $R$, well beyond the range of electron exchange. Therefore the dynamics of ultracold gases is dominated by their weak long-range (van der Waals) interaction varying as $R^{-6}$, which is isotropic for identical particles in free space with spherical symmetry (like atoms, or molecules with vanishing total angular momentum). Once the particles are immersed in an external magnetic or electric field, their intrinsic properties (permanent magnetic or electric dipole moment) induce the anisotropy of their long-range interaction, which can now vary as $R^{-3}$ and depends on the relative orientation of their molecular axis, or of their angular momentum \cite{lepers2013,zuchowski2013}. Manifestations of anisotropic interactions have already been observed experimentally with ultracold quantum degenerate gases of magnetic atoms \cite{stuhler2005,bismut2012,aikawa2012,lu2011}, and during ultracold collisions between KRb polar molecules (\textit{i.e.} possessing a permanent electric dipole moment in their own frame) \cite{ni2010,yan2013}. Such so-called ultracold dipolar gases are expected to reveal novel physical phenomena for instance in the context of quantum degeneracy where hamiltonians involved in condensed matter physics could be simulated with the opportunity for controlling the interaction between particles with external fields (see for instance the review papers of Refs.\cite{baranov2012,lahaye2009}). As for molecules, the recent review articles of Refs.\cite{quemener2012,stuhl2014} provide in-depth presentations of theory and experiments of collisions and reactions with ultracold molecules, emphasizing on their implications in the development of the new research area of ultracold chemistry dominated by quantum mechanical effects \cite{nesbitt2012}.

Despite amazing experimental progress, the main challenge for experimentalists dealing with ultracold molecules is still their formation as a gaseous sample with sufficient number density and with a good control of their internal state. An overview of the various methodologies to create ultracold neutral molecules and of their potential opportunities and applications is available in several review articles \cite{carr2009,dulieu2009,jin2012}, and we will not cover them here. In brief, there are two classes of approaches to obtain ultracold ground state molecules: (i) manipulating pre-existing polar diatomic or polyatomic molecules with external magnetic or electric fields to design slow molecular beams \cite{vandemeerakker2012,hutzler2012,narevicius2012}, and for some specific polar species, cooling diatomic molecules with laser \cite{dirosa2004,hummon2013,zhelyazkova2014,barry2014,kobayashi2014}; (ii) associating a pair of ultracold atoms into an ultracold molecule using laser photoassociation (PA) toward an electronic excited state followed by radiative emission (RE) to the ground state \cite{jones2006,ulmanis2012}, or magnetoasssociation (MA) \cite{kohler2006} in a weakly bound level of the electronic ground state manifold via magnetically tunable Feshbach resonances \cite{chin2010}, with a subsequent stimulated radiative transfer (SRT) process to populate the lowest molecular bound level of the ground state.

We focus for the rest of this paper on the latter option (MA+SRT), which has been successfully demonstrated experimentally in a still limited number of cases with homonuclear molecules like Cs$_2$ \cite{danzl2010} and Rb$_2$ \cite{lang2008}, and heteronuclear molecules like KRb \cite{ni2008} and RbCs \cite{takekoshi2014,gregory2015}. This is a quite general method for the class of alkali-metal diatomic molecules which all possess a wealth of Feshbach resonances (see Ref.\cite{patel2014} for KCs isotopologues). The population transfer toward the lowest bound level of the ground state is achieved via the well-known coherent process of stimulated Raman Adiabatic Passage (STIRAP, \cite{bergmann1998,vitanov2001,koch2012}). The efficiency of the transfer relies on the identification of a pair of so-called pump and dump electric dipole allowed transitions with comparable Rabi frequencies, and thus on the detailed knowledge of the spectroscopy of the molecule of interest. In a previous paper \cite{borsalino2014}, hereafter referred to as paper I, we modeled the STIRAP approach for the bosonic and fermionic KRb molecules. Using up-to-date spectroscopic data, we analyzed the efficiency of several transition schemes over the entire range of accessible laser frequencies determined by the excited electronic states, and we confirmed the suitability of the experimentally chosen scheme for KRb. We also investigated STIRAP efficiency on a limited range of frequencies for RbCs \cite{debatin2011,takekoshi2014}. 

Our study concerns the formation of ultracold bosonic $^{39}$K$^{133}$Cs and fermionic $^{40}$K$^{133}$Cs polar molecules in their absolute ground state by STIRAP, which has not yet been achieved experimentally. The present investigation aims at guiding ongoing experiments in the choice of their laser set-up to implement the STIRAP scheme. Its main outcome is that the most efficient STIRAP schemes do not generally correspond to the intuitive picture delivered by the Franck-Condon principle describing the strongest molecular transitions in terms limited to the spatial overlap of vibrational wave functions. The KCs species exhibits several specific properties which contrast with the KRb one studied in paper I. It possesses an intrinsic permanent electric dipole moment (PEDM) of 1.92~D almost 4 times larger than the KRb one \cite{aymar2005}. The potential well depth of the KCs molecular ground state being larger than the Cs$_2$ one, but smaller than the K$_2$ one, the KCs molecule is stable in its lowest ground state level against ultracold collisions with surrounding Cs atoms and other KCs molecules \cite{zuchowski2010}, but not with surrounding K atoms. These features will represent a decisive advantage for their further manipulation to create for instance a quantum degenerate molecular gas without the request of trapping molecules inside an optical lattice. Moreover, the KCs spectroscopy has been already quite well investigated experimentally for the X$^1\Sigma^+$ ground state and the lowest triplet state a$^3\Sigma^+$ \cite{ferber2009,ferber2013}, and for several excited states \cite{tamanis2010,kruzins2010,busevica2011,birzniece2012,klincare2012,kruzins2013}. The spectrum of KCs Feshbach resonances based on these results has also been modeled in detail \cite{patel2014}. 

The present paper is organized as follows. We first recall in Section \ref{sec:model} the basic principle of STIRAP, and we characterize the initial, and final molecular electronic states chosen for the implementation of STIRAP in $^{39}$KCs and in $^{40}$KCs (the index for the Cs mass will be omitted in the rest of the paper). The choice of the intermediate state for STIRAP is discussed in Section \ref{sec:interm}, emphasizing on the necessary knowledge of the requested molecular structure data, \textit{i.e.} potential energy curves (PECs), transition electric dipole moments (TEDMs) and spin-orbit couplings (SOCs). Three transition paths for STIRAP in KCs relying on different kinds of couplings between molecular levels are identified and their efficiencies are compared among each other (Section \ref{sec:results}). Experimental prospects are discussed in Section \ref{sec:prospects} in the perspective of the extension of such studies to other alkali-metal polar diatomic species.

When appropriate, the atomic unit of length ($1\,a_0= 0.052 917 721 092$~nm) and of dipole moment (1~a.u.$\equiv ea_0=$ 2.541 580 59~D) will be used.

\section{Model for STIRAP with KC\lowercase{s} molecules}
\label{sec:model}

The STIRAP principle has been proposed by Bergmann \textit{et al.}\cite{bergmann1998,vitanov2001} and further discussed for instance in Ref.\cite{koch2012} in the context of coherent photoassociation of ultracold atoms. We also presented a summary in paper I and we only recall here a few aspects which are relevant for the present study.

The central idea of STIRAP is to adiabatically transfer the population of a quantum system from an initial state \ket{i} to a well-defined final state \ket{g} via an intermediate excited state \ket{e}, in such a way that the \ket{e} state is actually not populated. This is achieved by cleverly shaping two laser pulses overlapping in time inducing  the \ket{i} $\rightarrow$ \ket{e} (pump) transition and the \ket{e} $\rightarrow$ \ket{g} (dump) transition. The Hamiltonian of the system dressed by the pulses admits a "`dark"' eigenstate which cannot radiatively decay and insures the transfer from \ket{i} to \ket{g} without loss of population. One can show that the optimal efficiency of the transfer is reached when one finds a level \ket{e} such as the amplitude of the time-dependent Rabi frequencies
\begin{equation}
    \overline{\Omega}_{ei} = \elem{e}{\vec{d} \cdot \vec{E}_{\mathrm{pump}}}{i} \, / \, \hbar \,\mathrm{;}\, \overline{\Omega}_{ge} = \elem{g}{\vec{d} \cdot \vec{E}_{\mathrm{dump}}}{e}\, / \, \hbar
\label{eq:STIRAP-1}
\end{equation}
for the pump and dump transitions are equal. In Eq.(\ref{eq:STIRAP-1}) $\vec{d}(R)$ is the electronic transition dipole moment function (TEDM), and $E_{\mathrm{pump}}$ (resp. $E_{\mathrm{dump}}$) is the amplitude of the laser field driving the pump (resp. dump) transition, associated to an intensity $I_{\mathrm{pump}}$ (resp $I_{\mathrm{dump}}$). This equality is achieved either with equal TEDM matrix elements for the pump and dump transitions assuming equal intensities for both pulses which is often convenient in experimental setups, or by slight adjustments of the laser intensities within the experimental feasibility to approach the strict equality of the Rabi frequencies.

Here \ket{g} is the lowest level $v_X=0$ of the X$^1\Sigma^+$ electronic ground state of KCs, and \ket{i} is a weakly-bound level of the X$^1\Sigma^+$ and a$^3\Sigma^+$ state manifolds (hereafter referred to as the X and a states) coupled by the hyperfine interaction. Figures~\ref{fig:PECs-0} and \ref{fig:PECs-1} display two possible choices of excited electronic state for the intermediate state \ket{e} that are discussed below.

The weakly-bound initial state \ket{i} results from the magnetoassociation of a pair of ultracold $^{39}$K (or $^{40}$K) and Cs atoms by tuning an external magnetic field onto a Feshbach resonance of the pair. Such a so-called Feshbach molecule is populated in a high-lying rovibrational level with a combination of triplet ($S=1$) and singlet ($S=0$) characters. The mixing coefficients depend on the choice of the Feshbach resonance, experimentally investigated in Ref.\cite{ferber2013} and accurately modeled in Ref.\cite{patel2014}. For a molecular system, there is an additional requirement for STIRAP to work, that the radial wavefunctions of the \ket{i} and \ket{e} on one hand, and of the \ket{e} and \ket{g} levels on the other hand should overlap each other in a region where the TEDMs are not vanishingly small. As it can be seen from Figs.\ref{fig:PECs-0}a and \ref{fig:PECs-1}a, this is well achieved if the \ket{i} levels contains a significant component on the a$^3\Sigma^+$ state which has an inner classical turning point in the suitable range of $R$. Just like in paper I, the first hypothesis of our model is to choose \ket{i} with a pure triplet character, so that one can describe it with a radial wavefunction belonging to the single a$^3\Sigma^+$ PEC with approximately the same binding energy as the one of the Feshbach molecule bound level. This is the case of the uppermost a$^3\Sigma^+$ level assigned to $v_a=35$ with a binding energy of about 0.05~GHz at zero magnetic field,  which possesses a triplet character up to 95\% \cite{ferber2013}. Note, however, that a pump transition starting from a pure singlet level would, in principle, be possible (see the dashed arrows in Fig.~\ref{fig:PECs-0}a) but is not discussed further here. 

Such an assumption is reasonable, as even if there is a significant singlet component in the chosen level, the nodal structure of the triplet component of the radial wavefunction will not be affected in the region of the inner turning point of the a$^3\Sigma^+$ PEC, while its amplitude may be changed. Therefore the matrix elements in Eq.~(\ref{eq:STIRAP-1}) would be affected only through a global scaling factor. An example of such a wave function concerns a level resulting from the mixture of two  a$^3\Sigma^+$ levels ($v_a = 32$ and $v_a = 33$) and two singlet levels ($v_X = 102$ and $v_a = 103$) (see Fig.6 of Ref.\cite{ferber2013}). In a recent proposal, Klincare \textit{et al.} \cite{klincare2012} proposed a STIRAP implementation based on a pump transition mainly acting around the outer turning point of the X PEC, using a similar hypothesis of a pure singlet weakly-bound level as the \ket{i} state. 

\section{The choice of the intermediate STIRAP levels in KCs}
\label{sec:interm}

\begin{figure*}
      \centering
      \includegraphics[scale = 0.6]{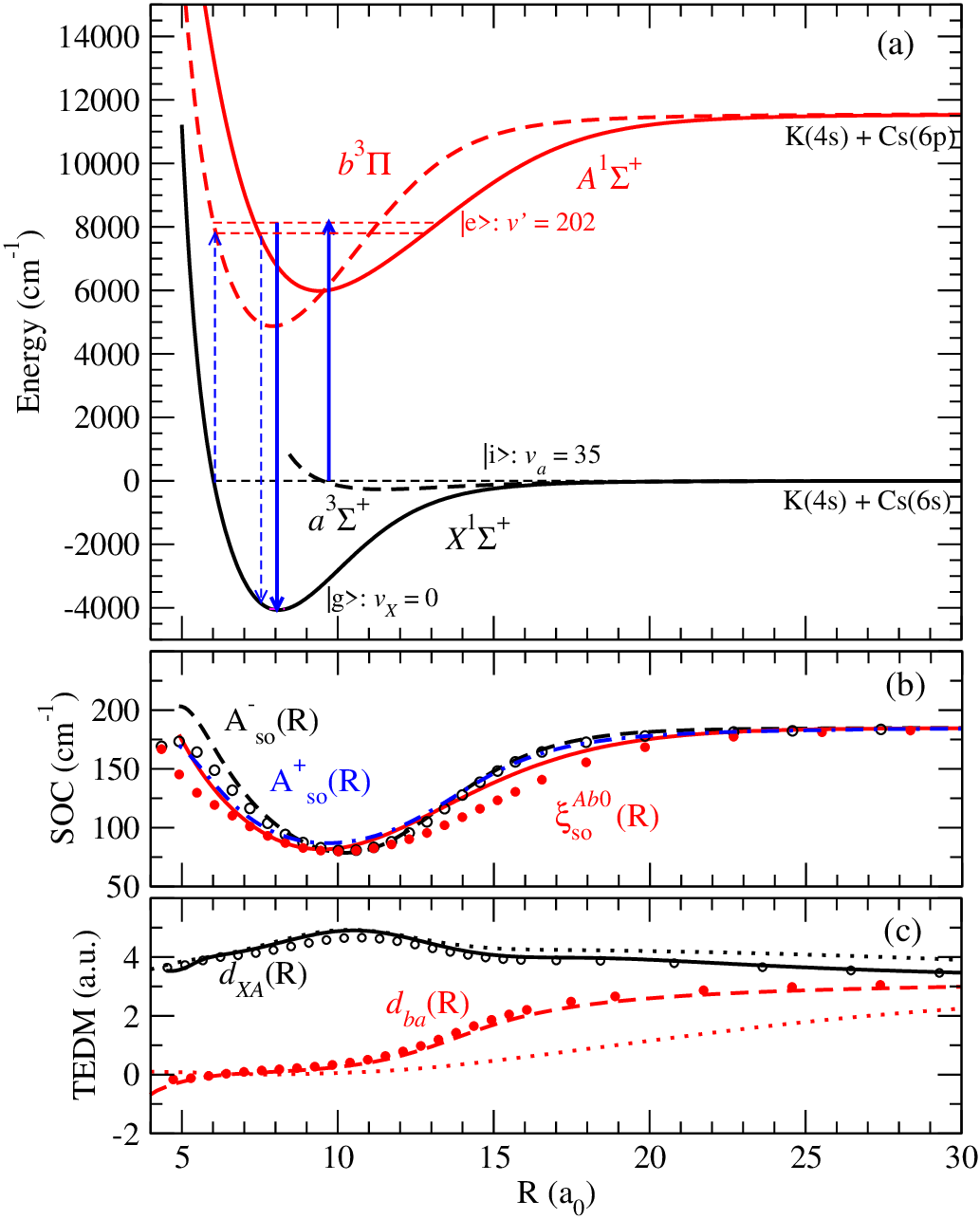}
      \caption{(a) Potential energy curves of KCs involved in the proposed STIRAP scheme (schematized with thick arrows) based on the \omO{} symmetry, and the related vibrational levels involved (see text). The possibility to initiate the pump transition from a pure singlet level is illustrated with the dashed arrows. (b) Diagonal $A^+_{\mathrm{so}}$ (blue dot-dashed line), $A^-_{\mathrm{so}}$ (red dashed line), and off-diagonal $\xi^{Ab0}_{\mathrm{so}}(R)$ (solid line) spin-orbit coupling (SOC) matrix elements coupling the A$^1\Sigma^+$ and b$^3\Pi$ excited molecular states used in the present work \cite{tamanis2010,kruzins2010}. The SOCs calculated in Ref.\cite{kim2009} are also displayed (open and closed circles). (c) Computed transition electric dipole moments (TEDMs) of KCs for the transition between the X and A singlet states (solid line), and between the a and b triplet states (dashed line), used in the present work. The same quantities for KRb used in Ref.\cite{borsalino2014} are displayed for comparison purpose (dotted lines). The TEDMs calculated in Ref.\cite{kim2009} are also displayed (open and closed circles).}
\label{fig:PECs-0}
\end{figure*}

\begin{figure*}
      \centering
      \includegraphics[scale = 0.6]{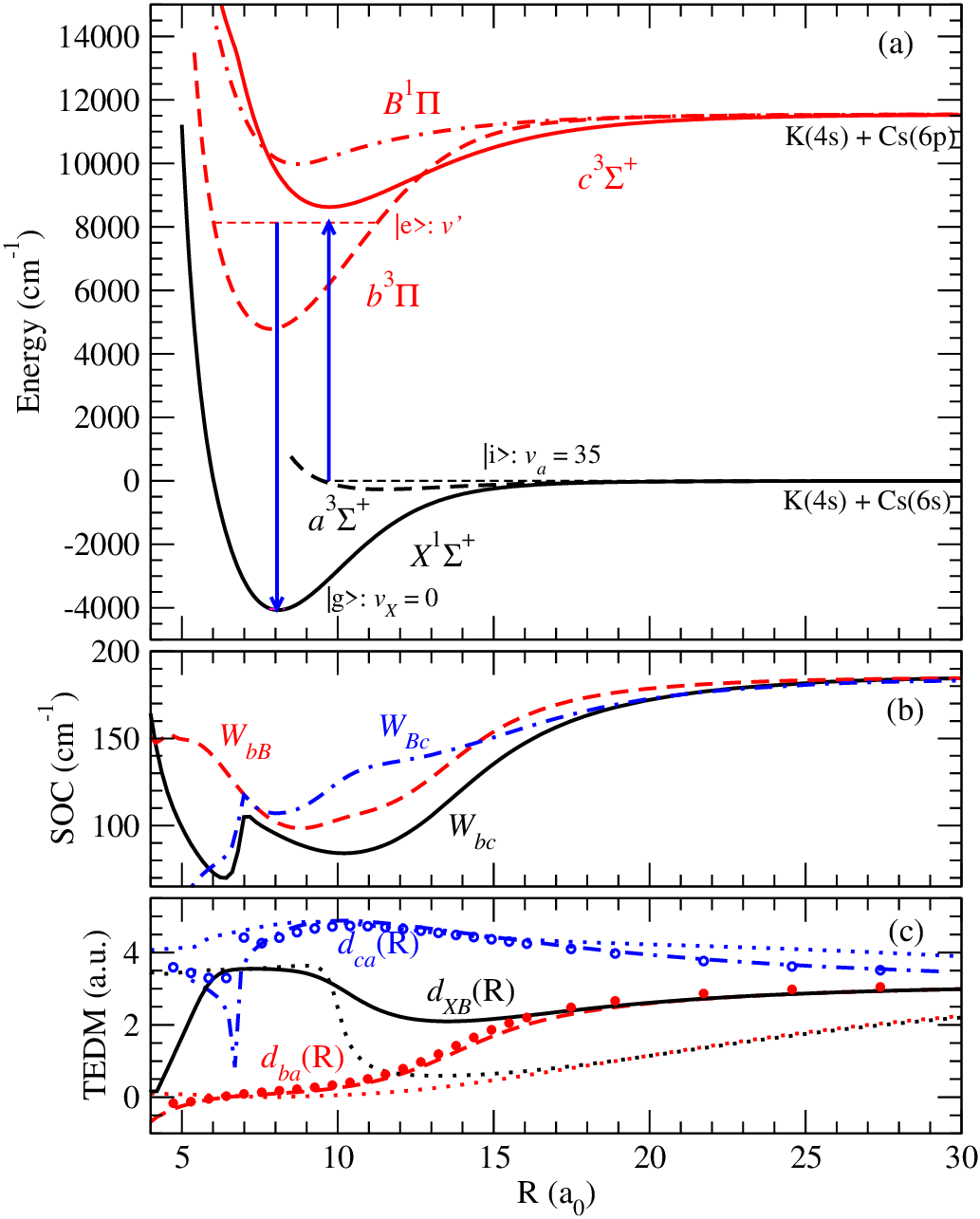}
      \caption{(a) Potential energy curves of KCs involved in the proposed STIRAP scheme (schematized with thick arrows) based on the \omI{} symmetry, and the related vibrational levels involved (see text). (b) Off-diagonal $W_{bc}(R)$ (solid line), $W_{bB}(R)$ (dashed line), and $W_{Bc}(R)$ (dashed-dotted line) spin-orbit coupling (SOC) matrix elements coupling the b$^3\Pi$, c$^3\Sigma^+$ and B$^1\Pi$ excited molecular states used in the present work \cite{kim2009}.(c) Computed transition electric dipole moments (TEDM) of KCs for the transition between the X and A singlet states (solid line), between the a and b triplet states (dashed lines), and between the a and c triplet states (dashed-dotted lines), used in the present work. The same quantities for KRb used in Ref.\cite{borsalino2014} are displayed for comparison purpose (dotted lines). The TEDMs calculated in Ref.\cite{kim2009} are also displayed (open and closed circles).}
\label{fig:PECs-1}
\end{figure*}

The \ket{e} state must be optically coupled to both  \ket{i} $\equiv$\ket{\wwa \: v_a = 35} and \ket{g} $\equiv$ \ket{\wwX \: v_X = 0} and thus must exhibit favorable transition dipole moments and good spatial overlaps with \ket{i} and \ket{g} vibrational wavefunctions. As in paper I, such a mixed singlet/triplet character is offered by the excited electronic states converging to the K($4s$) + Cs($6p$) dissociation limit, namely \wwb{}, \wwA{}, \wwc{} and \wwB{}, hereafter referred to as b, A, c and B states respectively (see Figs.\ref{fig:PECs-0}a and \ref{fig:PECs-1}a). These states are significantly affected by the spin-orbit (SO) interaction, resulting in coupled states with both spin characters, labeled with the Hund's case $(c)$ quantum number $\Omega=0^+, 1$ for the projection of the total electronic angular momentum on the molecular axis. In order to provide predictions to the experimentalists, it is crucial to rely on all available spectroscopic information about these states, so that the corresponding PECs are built piecewise, combining spectroscopic and quantum chemistry determinations. Note that the implementation proposed in Ref.\cite{klincare2012} relies on a higher electronic excited state, the (4)$^1\Sigma^+$ state correlated to the K($4s$)+Cs($5d$) dissociation limit, perturbed by neighboring triplet states.

Unlike the KRb case, our study of the $\Omega = 0^+$ symmetry has been greatly facilitated by the extensive spectroscopic study from  Refs.\cite{kruzins2010,tamanis2010,kruzins2013}, providing the relevant PECs (Fig.~\ref{fig:PECs-0}a) and $R$-dependent spin-orbit couplings (SOC) (Fig.~\ref{fig:PECs-0}b). Following these authors, we use a four-coupled-channel model which accounts for the dominant SO interaction between the \wwA{} state and the $\Omega = 0^+$ component b$_0$ of the \wwb{} state, as well as the interaction with other molecular states responsible for the asymmetric splitting between the $\Omega=0,1,2$ components of the b$^3\Pi$ state. In addition it includes the weak rotational interactions with the other $\Omega=1, 2$ components b$_1$ and b$_2$ of the \wwb{} state, scaling with $\beta = \hbar^2/(2\mu R^2)$ where $\mu$ is the KCs reduced mass. The resulting $R$-dependent potential energy matrix is expressed, for a given total angular momentum $J$ (including the total electronic angular momentum and the rotation of the molecule, but not the nuclear spins, and $X=J(J+1)$), as the sum of Born-Oppenheimer (BO) PEC matrix $V_{\mathrm{BO}}$ and SOC matrix $W_{\mathrm{so}}^{(0^+)}$
{\footnotesize
\begin{eqnarray}
\label{eq:WSO-0}
V_{\mathrm{BO}}+W_{\mathrm{so}}^{(0^+)} = \\ \nonumber
\bordermatrix{~ & \ket{b2}            &\ket{b1}               & \ket{b0}          & \ket{A }  \cr
& V'_{b2}(R)+A_{\mathrm{so}}^+(R)&-\beta(1-\gamma_b)\sqrt{2(X-2)}& 0 & 0         \cr
& -\beta(1-\gamma_b)\sqrt{2(X-2)}     &V'_{b1}(R) &-\beta(1-\gamma_b)\sqrt{2X} &-\beta\zeta_{Ab1}\sqrt{2X}\cr
& 0  &-\beta(1-\gamma_b)\sqrt{2X}   &V'_{b0}(R)-A_{\mathrm{so}}^-(R)&-\sqrt{2}\xi^{Ab0}_{\mathrm{so}}(R)\cr
& 0                   &-\beta\zeta_{Ab1}\sqrt{2X}&-\sqrt{2}\xi^{Ab0}_{\mathrm{so}}(R)& V'_A(R)  \cr
                   }
\end{eqnarray}
}
For compactness purpose we used the notations for the potential energies including the centrifugal term: $V'_{b2}(R)=V_b(R)+\beta (1-\gamma_b) (X-2)$, $V'_{b1}(R)=V_b(R)+\beta(1-\gamma_b)(X+2)$, $V'_{b0}(R)=V_b(R)+\beta(1-\gamma_b)(X+2)$, and $V'_A(R)=V_A(R)+\beta(1-\gamma_b)(X+2)$. This matrix reduces to a $3\times 3$ form if $J=1$. Note that, strictly speaking, the labels "`so"' and $0^+$ in $W_{\mathrm{so}}^{(0^+)}(R)$ are approximate, referring to the dominant SO interaction. We will keep them in the following, for convenience.

For the bound level calculations, we extend the experimental PECS of the A and b states at large distances with a $C_n / R^n$ ($n=6,8$) expansion using the $C_n$ coefficients from Ref.\cite{marinescu1999}. At large distances the functions $A_{\mathrm{so}}^+(R)$, $A_{\mathrm{so}}^-(R)$ and $\xi^{Ab0}_{\mathrm{so}}(R)$ reach the SO constant of the Cs atom, i.e. $\xi^{\mathrm{Cs}}_{\mathrm{so}} = \cm{184.68}$. The constants $\zeta_{Ab1}$ and $\gamma_b$ are empirically adjusted to take the relevant off-diagonal interaction into account. All those terms are obtained using the analytical formulas given in Refs.\cite{kruzins2010, tamanis2010,kruzins2013}. For completeness, we also report in Fig.~\ref{fig:PECs-0}b the SOC functions which were computed in Ref. \cite{kim2009} prior to the spectroscopic analysis of Ref.~\cite{ferber2013}, showing a remarkable agreement between the two determinations.

The amazing quality of the spectroscopic data for the \{b,A\} complex allows calculating the energies of the  $\Omega = 0^+$ levels with experimental precision. There are only a few observed levels assigned to levels with \wwb$_1$ and \wwb$_2$ character, so that the prediction of the energies of unobserved levels is not as accurate. However, as it has been experimentally demonstrated for RbCs \cite{docenko2010,takekoshi2014}, such a model indeed provides reliable information about these \wwb$_1$ levels.

For the $\Omega=1$ case, no full spectroscopic analysis for the b, c, and B coupled molecular states exists in the literature. Kim \textit{et al.} computed the relevant $R$-dependent SOC functions $W_{bc}$, $W_{bB}$, and $W_{Bc}$ with a quantum chemistry approach \cite{kim2009}, and we used  them in our model. The SO Hamiltonian matrix is expressed in a way similar to Eq.~(\ref{eq:WSO-0}) as
\begin{equation}
\label{eq:WSO-1}
V_{\mathrm{BO}}+W_{\mathrm{so}}^{(1)} = 
                                     \bordermatrix{      ~   & \ket{b}       & \ket{c}   & \ket{B}     \cr
                                                             & V'_b(R)             & W_{bc}(R) & -W_{bB}(R)  \cr
                                                             & W_{bc}(R)     & V'_c(R)      & W_{Bc}(R)   \cr
                                                             & -W_{bB}(R)    & W_{Bc}(R) & V'_B(R)           \cr
                                                     }
\end{equation}
with $V'_{\alpha}(R)=V_{\alpha}(R)+\beta X$ for $\alpha=$b, c, B. The coupling functions are reported in Fig.~\ref{fig:PECs-1}b, showing that they all converge toward $\xi^{\mathrm{Cs}}_{\mathrm{so}}$. For the bound level calculations, we used as above the experimental $b$ PEC \cite{tamanis2010,kruzins2013}. The c PEC is obtained from our own quantum chemistry calculations based on semi-empirical effective core potentials (ECP), following the procedure described in Refs.\cite{aymar2005,aymar2006a}, completed by the ECP parameters reported in Ref.~\cite{guerout2010}. The spectroscopy of the bottom of the B PEC has been achieved in Ref.\cite{birzniece2012}, which thus accounts for SOC in an effective way. Therefore we shifted it in energy before its connection to our own computed PEC curve in order to ensure that the three-coupled-channel calculation actually delivers the correct energies for the measured levels. As above, these PECs are connected at large distances to an asymptotic expansion using coefficients from Ref.\cite{marinescu1999}.

The treatment of the \omI{} complex is not as accurate as the \omO{} one, as there is no spectroscopic analysis available in the literature. However, the few spectroscopically observed vibrational levels at the bottom of the B state provide useful information for predicting a good STIRAP transfer using an \omI{} intermediate level.

In addition, TEDM functions connecting the X and a states with the A, b, B, and c states from our own quantum chemistry calculations are drawn in Fig.~\ref{fig:PECs-0}c and Fig.~\ref{fig:PECs-1}c. Note that as stated by Kim \textit{et al.} in their article \cite{kim2009}, both our own PECs and TEDMs are in excellent agreement with their results. It is also worthwhile to remark that the TEDMs are quite similar to the KRb ones, with an important difference however. The magnitude of $d_{ba}(R)$ is larger than its KRb counterpart \cite{borsalino2014} at the inner turning point of the a PEC: $d_{ba}$($R$=9.5a.u.) $\sim 0.21 ea_0$ in KCs, whereas in KRb $d_{ba}$($R$=9.3a.u.) $\sim 0.03 ea_0$. As discussed later, this order of magnitude difference will have important consequences for the experimental realization of the STIRAP approach based on $\Omega = 0^+$ states.

Our quantum chemistry data for PECs and TEDMs and the piecewise PECs elaborated above are provided in the Supplemental Material for convenience.

Finally, according to the authors, the semiempirical curves  and parameters derived in Refs.\cite{ferber2009,ferber2013,kruzins2010,tamanis2010,kruzins2013} are  correctly mass-invariant, so that it is possible to use them to model the levels of the other isotopologues $^{40}$KCs and $^{41}$KCs. This is partly verified for the \{b,A\} complex by the ability of the semiempirical curves to reproduce a few measured levels \cite{kruzins2010} of the $^{41}$K$^{133}$Cs molecule. For X and a PECs, the derived curves yield reliable scattering lengths for elastic collisions of each isotopic combination, thanks to the good quality of the long range part \cite{ferber2009, ferber2013,kruzins2013}.

\section{Three possible implementations of STIRAP in KCs}
\label{sec:results}

In order to evaluate the relevant transition matrix elements (TMEs) involved in the Rabi frequencies in Eq.(\ref{eq:STIRAP-1}), vibrational energies and radial wave functions are computed with the Mapped Fourier Grid Hamiltonian (MFGH) method \cite{kokoouline1999,kokoouline2000} as described in paper I. The Hamiltonian operator governing the nuclear motion is $\hat{H} = \hat{T} + \hat{V}_{\mathrm{BO}} + \hat{W}_{\mathrm{so}}^{(\Omega)}$, where $\hat{T}$ refers to the kinetic energy operator. It is represented as a $Nq \times Nq$ matrix where $N$ is the number of coupled channels and $q$ is the number of grid points for the $R$ coordinate determined such that the diagonalization yields eigenenergies reproducing the bound states energies at the experimental accuracy. The variable grid step at the heart of the MFGH method allows for accurately calculating bound levels with energies very close to the dissociation limit (and thus with a large vibrational amplitude) while limiting the number of grid points to about $q=590$. The resulting vibrational wave functions \ket{i} $\equiv$ \ket{\wwa \: v_a = 35}, and \ket{g} $\equiv$ \ket{\wwX \: v_X = 0}, are then used in our calculations. 

The vibrational levels resulting from the diagonalization are labeled with an index $v'_{\Omega}$ referring to the global numbering of the increasing eigenenergies. The corresponding radial wavefunctions $\left|\Omega; \,v'_{\Omega}\right\rangle$ are expressed as linear combinations of the $N$ coupled electronic states  
\begin{equation}
\left|\Omega; \,v'_{\Omega}\right\rangle = \sum_{\alpha=1}^N \frac{1}{R} \psi_{\alpha}^{\Omega v'_{\Omega}}(R) \left|\alpha\right\rangle .
\label{eq:wf}
\end{equation}
The weight $w_{\alpha}^{\Omega v'_{\Omega}}$ on each electronic state is defined by the squared radial components $\left|\psi_{\alpha}^{\Omega v'_{\Omega}}(R)\right|^2$ such as
\begin{equation}
\sum_{\alpha=1}^N w_{\alpha}^{\Omega v'_{\Omega}} \equiv \sum_{\alpha=1}^N \int_0^\infty \left|\psi_{\alpha}^{\Omega v'_{\Omega}}(R)\right|^2 dR   =1
\label{eq:norm}
\end{equation}
The TMEs for the pump (resp. dump) transition involve the vibrational functions $\varphi_a^{v_a}$ (resp.  $\varphi_X^{v_X}$) of the \wwa{} (resp. \wwX{}) state and the triplet part $\psi_{\alpha_t}^{\Omega v'_{\Omega}}(R)$ (resp. the singlet part $\psi_{\alpha_s}^{\Omega v'_{\Omega}}(R)$) of the coupled wave function of the intermediate level \ket{\Omega; v'_{\Omega}} (eq. (\ref{eq:wf}))
%
    \begin{eqnarray}
        {d}_{\alpha_t a}^{v'_{\Omega} v_a} &=& \left\langle \Omega; v'_{\Omega}\right| \hat{d}_{\alpha_t a} \left| a; v_a \right\rangle \nonumber \\
                                      &=& \int_0^\infty \psi_{\alpha_t}^{\Omega v'_{\Omega}}(R) \, d_{\alpha_t a}(R) \, \varphi_a^{v_a}(R) dR  \\
        {d}_{X \alpha_s}^{v_X v'_{\Omega}} &=& \left\langle X; v_X\right|\hat{d}_{X \alpha_s}\left| \Omega; v'_{\Omega} \right\rangle \nonumber \\
                                    &=& \int_0^\infty \varphi_X^{v_X}(R) \, d_{X \alpha_s}(R) \, \psi_{\alpha_s}^{\Omega v'_{\Omega}}(R) dR 
    \label{eq:d-matrix}
    \end{eqnarray}
%
The squared matrix elements $\left|{d}_{\alpha_t a}^{v'_{\Omega} v_a}\right|^2$ and $\left|{d}_{X \alpha_s}^{v_X v'_{\Omega}}\right|^2$ determine the efficiency of the STIRAP process and are systematically calculated in the following. Note that these TMEs will have to be multiplied by the appropriate H\"oln-London factors to take in account the experimentally chosen polarizations of the pump and dump lasers. 

In the next sections, graphs for TMEs will be drawn for the energy region where they are of comparable magnitude for the pump and dump transitions, for clarity. The full list of TMEs are given in the Supplementary Material attached to the present paper.

\subsection{STIRAP via the A -- b$_0$ spin-orbit coupled states}
\label{ssec:b0-A}

\begin{figure*}
\centering
\includegraphics[scale = 0.6]{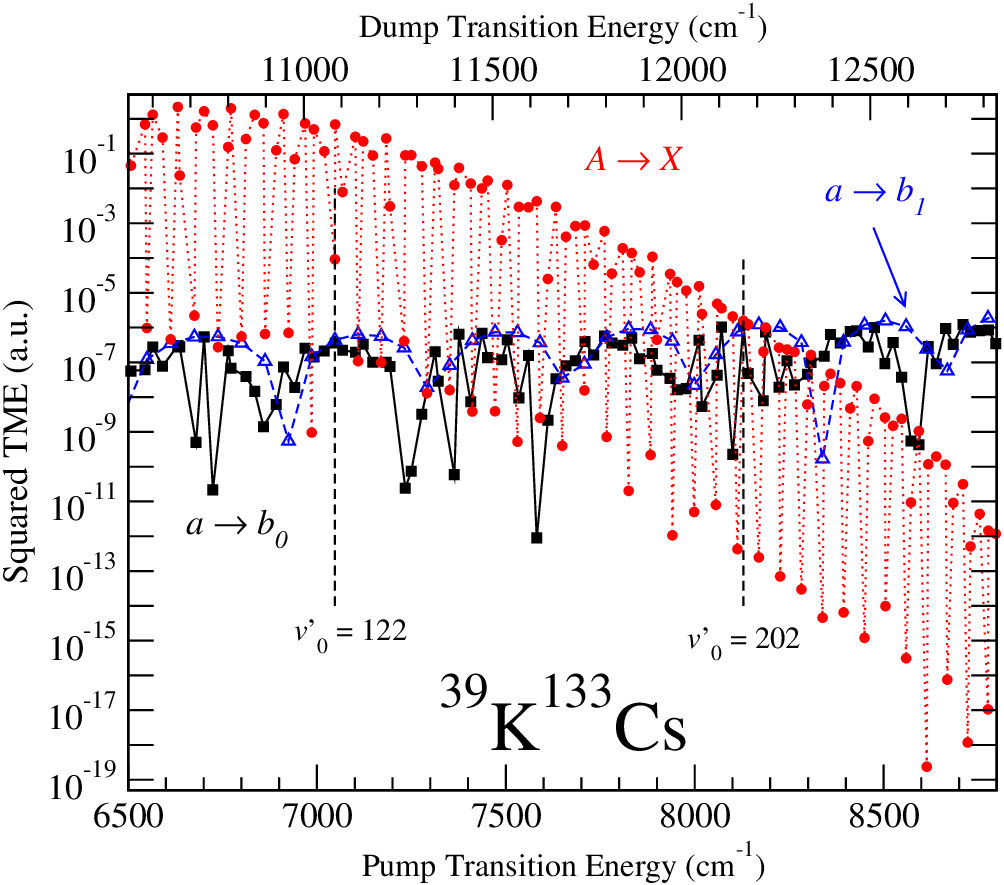}
\caption{Squared transition matrix elements (TME) in Eq. \ref{eq:d-matrix} for the pump transitions a$\rightarrow$b$_0$ (closed black squares) and a$\rightarrow$b$_1$ (open blue triangles), and for the dump transition A$\rightarrow$X (closed red circles) in the $^{39}$KCs isopotologue, for the $\Omega=0^+$ case. Rotational states are: $J=1$ for the intermediate levels, $J_X=0$ for the final $v_X=0$ level, and $J_a=0$ for the initial level $v_a=35$. The levels $v'_0$ for which the TME of the pump and dump transitions are equal are indicated.}
\label{fig:TME-0_39}
\end{figure*}

\begin{figure*}
      \centering
      \includegraphics[scale = 0.6]{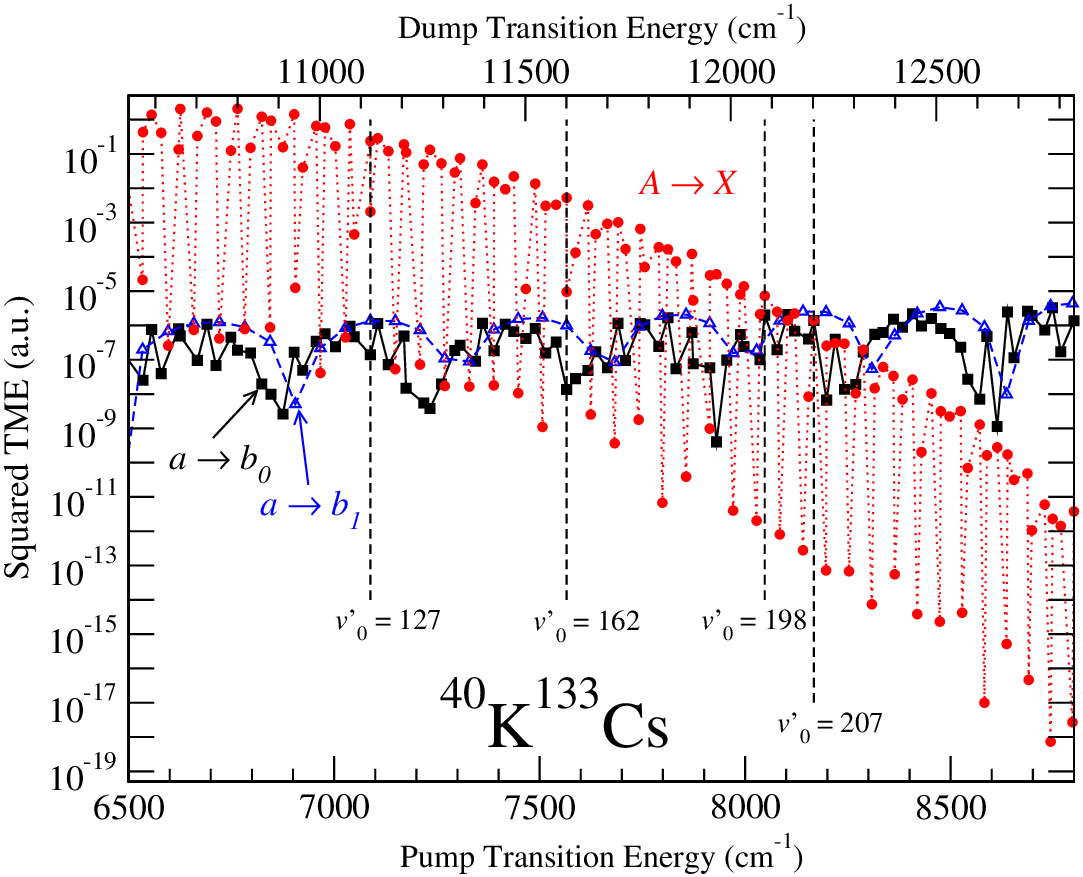}
      \caption{Same as Figure~\ref{fig:TME-0_39} for the $^{40}$KCs isopotologue. }
\label{fig:TME-0_40}
\end{figure*}

We considered the lowest allowed $J=1$ value for which the matrix of Eq.~(\ref{eq:WSO-0}) reduces to a $N=3$ dimension. We display in Fig.~\ref{fig:TME-0_39} for $^{39}$KCs and in Fig.~\ref{fig:TME-0_40} for $^{40}$KCs, the relevant TMEs ${d}_{b_0 a}^{v'_{0^+} v_a}$ (closed squares) and ${d}_{X A}^{v_X v'_{0^+}}$ (closed circles). They are extracted from the calculations above, involving the coupling matrix of Eq.(\ref{eq:WSO-0}).  As expected, their global behavior is very similar for the two isotopologues, but of course the recommended levels for the optimal transfer are slightly different, as summarized in Table \ref{tab:0+}. 

The data points for $\left|{d}_{X A}^{v_X v'_{0^+}}\right|^2$ (closed circles in Figs. \ref{fig:TME-0_39} and \ref{fig:TME-0_40}) are associated to \textit{all} eigenvectors yielded by the diagonalization, and of course related to the magnitude of their A component. The TMEs present strong variations associated to levels with main mixed A-b$_0$ character (upper zone of the data points, associated to states with strong A component, alternating with states with strong b$_0$ component), and to levels with main weight on the b$_1$ state (lower zone of the data, corresponding to states with very weak A component). The TMEs reported for $\left|{d}_{b_0 a}^{v'_{0^+} v_a}\right|^2$ (closed squares in Figs. \ref{fig:TME-0_39} and \ref{fig:TME-0_40}) correspond to states with either a main component on b$_0$ or on A, disregarding those with main b$_1$ character for clarity. Thus this data is complementary to the upper part of the data for $\left|{d}_{X A}^{v_X v'_{0^+}}\right|^2$.

Thus the optimal STIRAP region for equal pump and dump transitions, exemplified in the figures by the selected $v'_0=202$ level in $^{39}$KCs and by $v'_0=198$ and $v'_0=207$ in $^{40}$KCs, is located where the upper part of the A~$\rightarrow$ ~X data crosses the a~$\rightarrow$~b$_0$ data (with closed squares). Note that the picture is qualitatively similar than the one obtained in paper I for KRb, except that the b$_1$ and b$_2$ states were not included in the spin-orbit coupling matrix. However, the large $R$-dependent TEDM around the inner turning point of the a PEC in KCs compared to KRb taken in similar conditions, namely starting from the uppermost level $v=31$ of the a state (see Table \ref{tab:0+}) induces a much larger TME than in KRb, which makes this STIRAP scheme attractive for a future experimental implementation in KCs.

\begin{table*} 
\caption{Selected energies (in cm$^{-1}$) and matrix elements (in a.u.) of the pump and dump transitions (with energy $E_{pump}$ and $E_{dump}$) relevant for a STIRAP scheme based on an intermediate level belonging to the A -- b$_0$ spin-orbit coupled states  (with binding energy $E_{bind}$ relative to the $4s+6p$ dissociation limit) resulting from the spin-orbit coupling between A and b$_0$ levels (with $J=1$), starting from uppermost $v_a$=35 level (with $J=0$). H\"onl-London factors are not included. The weights on the various components of the intermediate level are also reported. The same results are recalled for $^{39}$K$^{87}$Rb \cite{borsalino2014}, with as the initial level the uppermost one $v_a=31$.}
{\footnotesize
\begin{tabular}{rrrrrrrrrr} \hline \hline
                 &$v'$&$E_{bind}$&$E_{pump}$&$E_{dump}$&$w_{b_0}$&$w_{b_1}$&$w_{A}$&$|d_{ab}|^2$&$|d_{AX}|^2$ \\ \hline
$^{39}$KCs       &202 &-3418.866 &8128.764  &12163.843  & 0.79    & 0.20(-6) &0.21   & 1.21(-6)   & 1.57(-6)      \\ 
$^{40}$KCs       &207 &-3380.794 &8166.841  &12202.246  & 0.31    & 0.13(-6) &0.69   & 1.40(-6)   & 1.40(-6)  \\ 
                 &198 &-3500.188 &8047.448  &12082.852  & 0.60    & 0.17(-6) &0.40   & 2.00(-6)   & 7.36(-6)  \\ 
$^{39}$K$^{87}$Rb&103 &-3450.1   &9287.297  &13467.396  & 0.09    &    -     &0.91   & 9.74(-9)   & 2.0 (-8)  \\
\hline \hline
\end{tabular}
}
\label{tab:0+}
\end{table*}

\subsection{STIRAP via the A -- b$_1$ rotationnally coupled states}
\label{ssec:b1-A}

The relevant TMEs are ${d}_{b_1 a}^{v'_{1} v_a}$ (open triangles in Figs. \ref{fig:TME-0_39} and \ref{fig:TME-0_40}) and ${d}_{X A}^{v_X v'_{0^+}}$ (the low set of closed circles in Figs. \ref{fig:TME-0_39} and \ref{fig:TME-0_40}). The data points correspond to levels with main b$_1$ character \textit{and} with the largest possible component on the A state (typically $\approx 10^{-4}$). Due to the weak rotational coupling (the constant $\beta$ in Eq.(\ref{eq:WSO-0}) amounts $\approx$0.04~cm$^{-1}$ around $R=$10~a.u.), vibrational levels of the unperturbed A and b$_1$ PECs must be quite close in energy to be effectively coupled. As the available KCs spectroscopic data is of good quality, we identified one level in $^{39}$KCs and two levels in $^{40}$KCs of main b$_1$ character with such characteristics (Table~\ref{tab:0+-1}). Despite small TMEs and quite unbalanced transition matrix elements for the pump and dump transitions, we predict a situation which is comparable to the one modeled and already observed in RbCs, which characteristics are recalled in Table \ref{tab:0+-1}. This is actually such a circumstance which recently allowed for an efficient STIRAP implementation to create a dense sample of ultracold $^{87}$RbCs molecules \cite{debatin2011,takekoshi2014}. This mechanism is expected to be even more favorable if a more deeply-bound level is chosen for \ket{i} (as done in RbCs \cite{debatin2011}) since the amplitude of its assumed pure triplet wave function around the a$^3\Sigma^+$ inner turning point grows up. 

\begin{table*} 
\caption{Selected energies (in cm$^{-1}$) and matrix elements (in a.u.) of the pump and dump transitions (with energy $E_{pump}$ and $E_{dump}$) relevant for a STIRAP scheme based on an intermediate level belonging to the A -- b$_1$ rotationnally coupled states (with binding energy $E_{bind}$ relative to the $4s+6p$ dissociation limit) resulting from the rotational coupling between A and b$_1$ levels (with $J=1$), starting from $v_a$=35 with $J=0$. H\"onl-London factors are not included. The weights on the various components of the intermediate level are also reported. The results for the same mechanism experimentally implemented in $^{87}$RbCs is recalled \cite{debatin2011}, starting from the 6$^{th}$downward the dissociation limit, corresponding to $v_a=42$. }
{\footnotesize
\begin{tabular}{rrrrrrrrrr}\hline \hline
           &$v'$&$E_{bind}$&$E_{pump}$&$E_{dump}$&$w_{b_1}$&$w_{b_0}$&$w_{A}$&$|d_{ab}|^2$&$|d_{AX}|^2$ \\ 
$^{39}$KCs   &122 &-4500.405 &7047.225  &11082.304 &0.99987  &0.00003  &0.00010& 4.42(-7)  & 9.19(-5)  \\ 
$^{40}$KCs   &127 &-4459.692 &7087.944  &11123.348 &0.99138  &0.00191  &0.00671& 1.37(-6)  & 2.05(-3)  \\ 
             &162 &-3982.269 &7565.367  &11600.771 &0.99826  &0.00055  &0.00118& 1.01(-6)  & 9.24(-6)  \\ 
$^{87}$RbCs  & 68 &-5124.586 &6423.042  &10234.613 &0.99603  &0.00083  &0.00314& 9.49(-7)  & 1.71(-4)  \\ 
\hline  \hline
\end{tabular}
}
\label{tab:0+-1}
\end{table*}

\subsection{STIRAP via the B--b--c spin-orbit coupled states}
\label{ssec:B-c}

\begin{figure*}
      \centering
      \includegraphics[scale = 0.6]{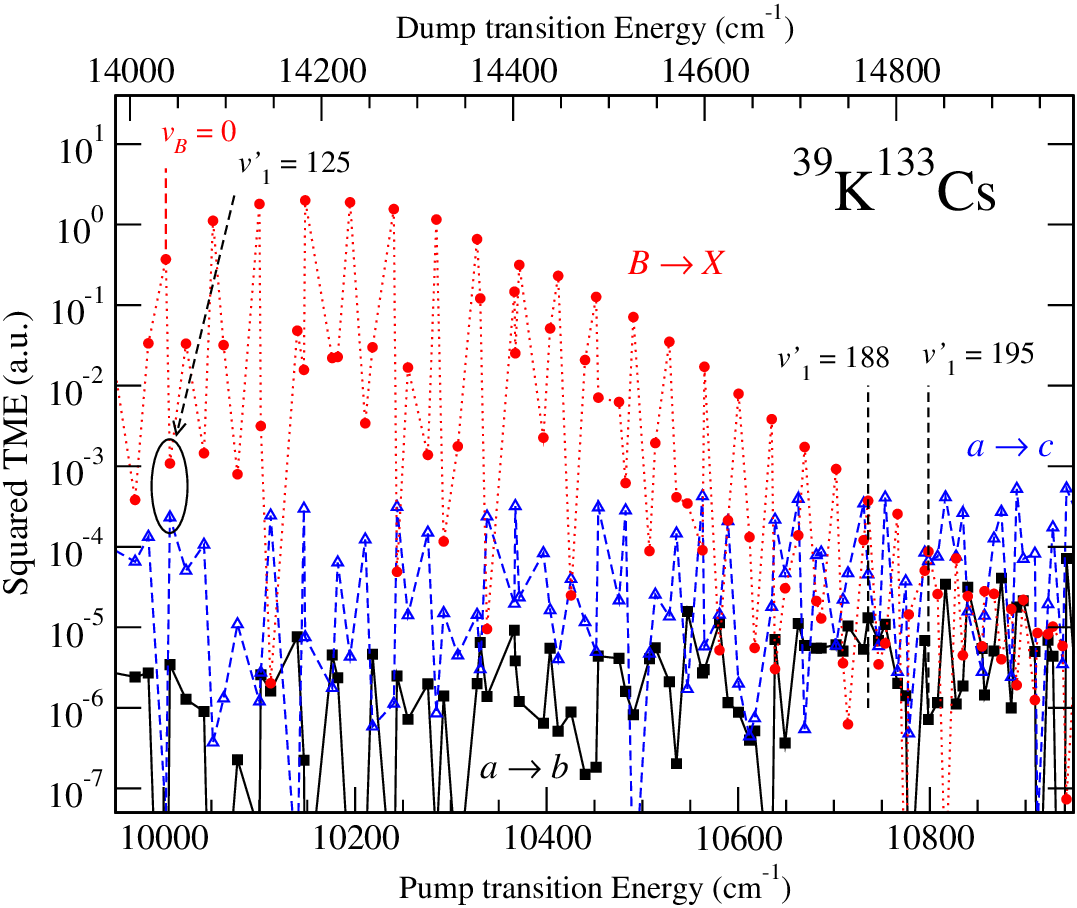}
      \caption{Squared transition matrix elements (TME) in Eq. \ref{eq:d-matrix} for the pump transitions a$\rightarrow$b (closed black squares) and a$\rightarrow$c (open blue triangles), and for the dump transition B$\rightarrow$X (closed red circles) in the $^{39}$KCs isopotologue, for the $\Omega=1$ case. Rotational states are: $J=1$ for the intermediate levels, $J_X=0$ for the final $v_X=0$ level, and $J_a=0$ for the initial level $v_a=35$. The levels $v'_1$ for which the TME of the pump and dump transitions are equal are indicated.}
\label{fig:TME-1}
\end{figure*}

The relevant TMEs ${d}_{b_1 a}^{v'_{1} v_a}$, ${d}_{c a}^{v'_{1} v_a}$, and ${d}_{X B}^{v_X v'_{1}}$ are extracted from the calculations involving the coupling matrix of Eq.(\ref{eq:WSO-1}). They are presented in Fig.~\ref{fig:TME-1} for $^{39}$KCs as a representative isotopologue, as the full spectroscopy of the $\Omega=1$ states is not yet available. The recommended levels for an optimal STIRAP implementation are displayed in Table~\ref{tab:1-Bc}. As in KRb \cite{borsalino2014}, due to the larger TEDM for the a~$\rightarrow$~c than for the a~$\rightarrow$~b transition, these levels are characterized by TMEs of comparable magnitude for the pump ~$\rightarrow$~c transition and the dump B~$\rightarrow$~X transition. Thus we qualify this case as being induced by the B--c spin-orbit coupling. But in contrast with KRb, it is likely to reach quite high-lying B vibrational levels ($v_B=$20, 23, while for KRb we had $v_B=$8) to ensure an optimal STIRAP. These levels are located at an energy corresponding to the quantum chemistry part of the B PEC, while its spectroscopic determination is yielded only up to the energy of $v_B=5$ in Ref.\cite{birzniece2012}. In the $v_B \leq 5$ energy range a couple of $\Omega=1$ levels with main b character are also expected to be interesting for STIRAP and are reported in Table \ref{tab:1-Bc}. For instance the $v' = 125$ level energy is predicted close to the location of $v_B=0$ which is well known experimentally. Due to their noticeable weight on the B state, such levels are most likely present in the recorded data of Ref.\cite{birzniece2012}, even if not yet assigned.

Finally, this B--b--c STIRAP option is based on TMEs which are larger than those for the A--b$_0$ and A--b$_1$ options by about two orders of magnitude only, in strong contrast with KRb where the difference was at least of four orders of magnitude. As anticipated above, this is due to the large a--b TEDM in KCs compared to the KRb one.

\begin{table*} 
\caption{Selected energies (in cm$^{-1}$) and matrix elements (in a.u.) of the pump (a~$\rightarrow$~c and a~$\rightarrow$~b) and dump (B~$\rightarrow$~X) transitions (with energy $E_{pump}$ and $E_{dump}$) relevant for a STIRAP scheme based on an intermediate level belonging to the B--b--c spin-orbit coupled states (with binding energy $E_{bind}$ relative to the $4s+6p$ dissociation limit) resulting from the spin-orbit coupling between the b, c, and B states (with $J=1$), starting from $v_a$=35 (with $J=0$). H\"onl-London factors are not included. The weights on the various components of the intermediate level are also reported. The binding energy of $v_B = 0$ is \cm{-1547.6}. The main $v_B$ or $v_b$ wavefunctions involved in the full $v'$ coupled vibrational wave functions are indicated.}
{\footnotesize
\begin{tabular}{rrrrrrrrrrr}\hline \hline
          &$v'$&$E_{bind}$&$E_{pump}$&$E_{dump}$&$w_{b}$&$w_{c}$&$w_{B}$&$|d_{ba}|^2$&$|d_{ca}|^2$&$|d_{XB}|^2$ \\ 
$^{39}$KCs&188       &-814.4 &10733.3  &14768.4  &0.193 & 0.275 & 0.532 & 1.32(-5) & 4.54(-5) & 3.72(-4)  \\ 
          &($v_B=20$)&       &         &         &      &       &       &          &          &           \\
          &195       &-751.5 &10796.2  &14831.2  &0.138 & 0.282 & 0.579 & 7.18(-7) & 6.64(-5) & 8.71(-5)  \\ 
          &($v_B=23$)&       &         &         &      &       &       &          &          &           \\
          &125       &-1543.3&10004.3  &14039.4  &0.630 & 0.357 & 0.013 & 3.44(-6) & 2.32(-4) & 1.08(-3) \\ 
          &($v_b=90$)&       &         &         &      &       &       &          &          &           \\
          &163       &-1067.5&10480.1  &14515.2  &0.528 & 0.440 & 0.032 & 1.60(-6) & 2.83(-4) & 6.17(-4)  \\ 
          &($v_b=100$)&       &         &         &      &       &       &          &          &           \\
\hline \hline
\end{tabular}
\label{tab:1-Bc}
}
\end{table*}

\section{Prospects for experimental implementation}
\label{sec:prospects}

The present investigation of the possible pathways for the formation of ultracold KCs molecules in their absolute ground state is a follow-up of our previous study on KRb. But the current situation on the experimental side is very different for the two molecules.

The formation of ultracold KRb molecules in their absolute ground state via a STIRAP scheme has been undoubtedly boosted in part by the wealth of spectroscopic data available for the B$^1\Pi$ state in the region of its PEC minimum, with an accurate modeling of the perturbations by the b$^3\Pi$ and c$^3\Sigma^+$ states induced by spin-orbit interaction. Our complete analysis in paper I confirmed the choice of the experimental groups for the implementation of STIRAP based on the B--b--c scheme. Indeed, the weak TEDM for the a$^3\Sigma^+ \rightarrow$ b$^3\Pi$ pump transition does not favor the implementation via the other A--b$_0$ scheme relying on the spin-orbit coupled A$^1\Sigma^+$ and b$^3\Pi$ states.

The situation is reversed for the KCs species which has not yet been observed in the ultracold regime. The spectroscopy of the coupled A$^1\Sigma^+$ and b$^3\Pi$ states is much better known than the one of the B$^1\Pi$, b$^3\Pi$ and c$^3\Sigma^+$ states. Moreover, the large $R$-dependent TEDM around the inner turning point of the a PEC in KCs compared to KRb induces a much larger TMEs for the A--b$_0$ and A--b$_1$ schemes in KCs, which makes this STIRAP scheme attractive for a future experimental implementation. Thus our study shows that there are more possible options than in KRb to implement efficient STIRAP in KCs, namely, using either the A--b$_0$ spin-orbit coupled states, the A--b$_1$ rotationnally coupled states, or the B--b--c spin-orbit coupled states. Note that the A--b$_1$ case was not analyzed in paper I, but it has been successfully implemented in the RbCs experiment of Ref.\cite{debatin2011,takekoshi2014}. The magnitudes of the relevant RbCs transition matrix elements are recalled in Table \ref{tab:0+-1}, which provide a reference for the experimental feasibility of such a scheme in the KCs case. In this respect the STIRAP scheme based on A--b$_1$ coupled states is particularly attractive: it relies on the accurately known spectroscopy of the A--b coupled states. Moreover, despite the very weak hyperfine structure expected for $\Omega=0^+$ levels, their coupling with closeby $\Omega=1$ levels identified in KCs, -which are expected to possess a large hyperfine structure- allows for controlling the hyperfine level during STIRAP, as achieved for RbCs. We also predict that the STIRAP scheme yielding the largest TMEs for both the pump and dump transitions is the one based on the B--b--c spin-orbit coupled states, as observed in KRb. However, its implementation would require further spectroscopic investigations, which may not be the priority of the interested experimental groups.

Generally, the presented study, just like the previous one on KRb, demonstrates that the choice of an efficient STIRAP scheme to create ultracold molecules in their absolute ground state level cannot be determined by invoking the Franck-Condon (FC) principle which only involves the spatial overlap of the vibrational wave functions. The variation of the relevant TEDMs along the internuclear distance plays a central role. Moreover a reasonable balance of the TMEs for the pump and dump transitions should be achieved, which generally corresponds to transitions departing from the most favorable ones identified by the FC principle. In this respect, the provided Supplementary material should be a great help for setting up an experiment. We anticipate that the present study comes at the appropriate time to guide future experiments aiming at creating ultracold samples of KCs molecules. The hyperfine structure not taken into account in this work will be modeled in an upcoming study and the STIRAP schemes will reexamined in this framework.

\section*{Acknowledgments}
This project is supported in part by the contract BLUESHIELD of Agence Nationale de la Recherche (ANR, contract ANR-14-CE34-0006). R.V. acknowledges partial support from ANR under the project COPOMOL (contract ANR-13-IS04-0004). Stimulating discussions with Hanns-Christoph Nägerl and Emil Kirilov are gratefully acknowledged.

\newpage
\section*{References}

\begin{thebibliography}{10}
\expandafter\ifx\csname url\endcsname\relax
  \def\url#1{{\tt #1}}\fi
\expandafter\ifx\csname urlprefix\endcsname\relax\def\urlprefix{URL }\fi
\providecommand{\eprint}[2][]{\url{#2}}

\bibitem{lepers2013}
Lepers M, Vexiau R, Aymar M, Bouloufa-Maafa N and Dulieu O 2013 {\em Phys. Rev.
  A\/} {\bf 88} 032709

\bibitem{zuchowski2013}
$\dot{Z}$uchowski, Kosicki M, Kodrycka M and Sold{\`a}n P 2013 {\em Phys. Rev.
  A\/} {\bf 87} 022706

\bibitem{stuhler2005}
Stuhler J, Griesmaier A, Koch T, Fattori M, Pfau T, Giovanazzi S, Pedri P and
  Santos L 2005 {\em Phys. Rev. Lett.\/} {\bf 95} 150406

\bibitem{bismut2012}
Bismut G, Laburthe-Tolra B, Mar\'echal E, Pedri P, Gorceix O and Vernac L 2012
  {\em Phys. Rev. Lett.\/} {\bf 109} 155302

\bibitem{aikawa2012}
Aikawa K, Frisch A, Mark M, Baier S, Rietzler A, Grimm R and Ferlaino F 2012
  {\em Phys. Rev. Lett.\/} {\bf 108} 210401

\bibitem{lu2011}
Lu M, Burdick N~Q, Youn S~H and Lev B~L 2011 {\em Phys. Rev. Lett.\/} {\bf 107}
  190401

\bibitem{ni2010}
Ni K~K, Ospelkaus S, Wang D, Qu\'em\'ener G, Neyenhuis B, de~Miranda M~H~G,
  Bohn J~L, Ye J and Jin D~S 2010 {\em Nature\/} {\bf 464} 1324

\bibitem{yan2013}
Yan B, Moses S~A, Gadway B, Covey J~P, Hazzard K~R~A, Rey A~M, Jin D~S and Ye J
  2013 {\em Nature\/} {\bf 501} 521

\bibitem{baranov2012}
Baranov M~A, Dalmonte M, Pupillo G and Zoller P 2012 {\em Chem. Rev.\/} {\bf
  112} 5012

\bibitem{lahaye2009}
Lahaye T, Menotti C, Santos L, Lewenstein M and Pfau T 2009 {\em Rep. Prog.
  Phys.\/} {\bf 72} 126401

\bibitem{quemener2012}
Qu{\'e}m{\'e}ner G and Julienne P~S 2012 {\em Chem. Rev.\/} {\bf 112} 4949

\bibitem{stuhl2014}
Stuhl B~K, Hummon M~T and Ye J 2014 {\em Annu. Rev. Phys. Chem.\/} {\bf 65} 501

\bibitem{nesbitt2012}
Nesbitt D~J 2012 {\em Chem. Rev.\/} {\bf 112} 5062

\bibitem{carr2009}
Carr L~D and Ye J 2009 {\em New J. Phys.\/} {\bf 11} 055009

\bibitem{dulieu2009}
Dulieu O and Gabbanini C 2009 {\em Rep. Prog. Phys.\/} {\bf 72} 086401

\bibitem{jin2012}
Jin D and Ye J 2012 {\em Chem. Rev.\/} {\bf 112} 4801

\bibitem{vandemeerakker2012}
van~de Meerakker S~Y~T, Bethlem H~L, Vanhaecke N and Meijer G 2012 {\em Chem.
  Rev.\/} {\bf 112} 4828

\bibitem{hutzler2012}
Hutzler N~R, Lu H~I and Doyle J~M 2012 {\em Chem. Rev.\/} {\bf 112} 4803

\bibitem{narevicius2012}
Narevicius E and Raizen M~G 2012 {\em Chem. Rev.\/} {\bf 112} 4879

\bibitem{dirosa2004}
Rosa M~D~D 2004 {\em Eur. Phys. J. D\/} {\bf 31} 395

\bibitem{hummon2013}
Hummon M~T, Yeo M, Stuhl B~K, Collopy A~L, Xia Y and Ye J 2013 {\em Phys. Rev.
  Lett.\/} {\bf 110} 143001

\bibitem{zhelyazkova2014}
Zhelyazkova V, Cournol A, Wall T~E, Matsushima A, Hudson J~J, Hinds E~A,
  Tarbutt M~R and Sauer B~E 2014 {\em Phys. Rev. A\/} {\bf 89} 053416

\bibitem{barry2014}
Barry J~F, McCarron D~J, Norrgard E~B, Steinecker M~H and DeMille D 2014 {\em
  Nature\/} {\bf 512} 286

\bibitem{kobayashi2014}
Kobayashi J; Aikawa K; Oasa~K I~S 2014 {\em Phys. Rev. A\/} {\bf 89} 021401

\bibitem{jones2006}
Jones K~M, Tiesinga E, Lett P~D and Julienne P~S 2006 {\em Rev. Mod. Phys.\/}
  {\bf 78} 483

\bibitem{ulmanis2012}
Ulmanis J, Deiglmayr J, Repp M, Wester R and Weidem\"uller M 2012 {\em Chem.
  Rev.\/} {\bf 112} 4890

\bibitem{kohler2006}
K\"ohler T, G\'{o}ral K and Julienne P~S 2006 {\em Rev. Mod. Phys.\/} {\bf 78}
  1311

\bibitem{chin2010}
Chin C, Grimm R, Julienne P and Tiesinga E 2010 {\em Rev. Mod. Phys.\/} {\bf
  82} 1225

\bibitem{danzl2010}
Danzl J~G, Mark M~J, Haller E, Gustavsson M, Hart R, Aldegunde J, Hutson J~M
  and N\"agerl H~C 2010 {\em Nature Phys.\/} {\bf 6} 265

\bibitem{lang2008}
Lang F, van~der Straten P, Brandst\"{a}tter B, Thalhammer G, Winkler K,
  Julienne P~S, Grimm R and Denschlag J~H 2008 {\em Nature Physics\/} {\bf 4}
  223

\bibitem{ni2008}
Ni K~K, Ospelkaus S, de~Miranda M~H~G, Peer A, Neyenhuis B, Zirbel J~J,
  Kotochigova S, Julienne P~S, Jin D~S and Ye J 2008 {\em Science\/} {\bf 322}
  231

\bibitem{takekoshi2014}
Takekoshi T, Reichs\"ollner L, Schindewolf A, Hutson J~M, Sueur C~R~L, Dulieu
  O, Ferlaino F, Grimm R and N\"agerl H~C 2014 {\em Phys. Rev. Lett.\/} {\bf
  113}

\bibitem{gregory2015}
Gregory P~D, Molony P~K, Kumar A, Ji Z, Lu B, Marchant A~L, Cornish S~L 2015 {\em  	arXiv:1411.7951 [physics.atom-ph]\/}

\bibitem{patel2014}
Patel H~J, Blackley C~L, Cornish S~L and Hutson J~M 2014 {\em Phys. Rev. A\/}
  {\bf 90} 032716

\bibitem{bergmann1998}
Bergmann K, Theuer H and Shore B~W 1998 {\em Rev. Mod. Phys.\/} {\bf 70} 1003

\bibitem{vitanov2001}
Vitanov N~V, Fleischhauer M, Shore B~W and Bergemann W 2001 {\em Adv. At. Mol.
  Opt. Phys.\/} {\bf 46} 55

\bibitem{koch2012}
Koch C~P and Shapiro M 2012 {\em Chem. Rev.\/} {\bf 112} 4928

\bibitem{borsalino2014}
Borsalino D, Londo\~no Flor\`ez B, Vexiau R, Dulieu O, Bouloufa-Maafa N and
  Luc-Koenig E 2014 {\em Phys. Rev. A\/} {\bf 90} 033413

\bibitem{debatin2011}
Debatin M, Takekoshi T, Rameshan R, Reichsoellner L, Ferlaino F, Grimm R,
  Vexiau R, Bouloufa N, Dulieu O and Naegerl H~C 2011 {\em Phys. Chem. Chem.
  Phys.\/} {\bf 13} 18926

\bibitem{aymar2005}
Aymar M and Dulieu O 2005 {\em J. Chem. Phys.\/} {\bf 122} 204302

\bibitem{zuchowski2010}
\.Zuchowski P~S, Aldegunde J and Hutson J~M 2010 {\em Phys. Rev. A\/} {\bf 81}
  060703(R)

\bibitem{ferber2009}
{R Ferber}, {I Klincare}, {O Nikolayeva}, {M Tamanis}, {H Kn\"ockel}, {E
  Tiemann} and {A Pashov} 2009 {\em Phys. Rev. A\/} {\bf 80} 062501

\bibitem{ferber2013}
Ferber R, Nikolayeva O, Tamanis M, Kn\"ockel H and Tiemann E 2013 {\em Phys.
  Rev. A\/} {\bf 88} 012516

\bibitem{tamanis2010}
Tamanis M, Klincare I, Kruzins A, Nikolayeva O, Ferber R, Pazyuk E~A and
  Stolyarov A~V 2010 {\em Phys. Rev. A\/} {\bf 82} 032506

\bibitem{kruzins2010}
Kruzins A, Klincare I, Nikolayeva O, Tamanis M, Ferber R, Pazyuk E~A and
  Stolyarov A~V 2010 {\em Phys. Rev. A\/} {\bf 81} 042509

\bibitem{busevica2011}
{L Busevica}, {I Klincare}, {O Nikolayeva}, {M Tamanis}, {R Ferber}, {V V
  Meshkov}, {E A Pazyuk} and {A V Stolyarov} 2011 {\em J. Chem. Phys.\/} {\bf
  134} 104307

\bibitem{birzniece2012}
{I Birzniece}, {O Nikolayeva}, {M Tamanis} and {R Ferber} 2012 {\em J. Chem.
  Phys.\/} {\bf 136} 064304

\bibitem{klincare2012}
Klincare I, Nikolayeva O, Tamanis M, Ferber R, Pazyuk E~A and Stolyarov A~V
  2012 {\em Phys. Rev. A\/} {\bf 85} 062520

\bibitem{kruzins2013}
Kruzins A, Klincare I, Nikolayeva O, Tamanis M, Ferber R, Pazyuk E~A and
  Stolyarov A~V 2013 {\em J. Chem. Phys.\/} {\bf 139} 244301

\bibitem{marinescu1999}
Marinescu M and Sadeghpour H~R 1999 {\em Phys. Rev. A\/} {\bf 59} 390

\bibitem{kim2009}
Kim J, Lee Y and Stolyarov A 2009 {\em J. Mol. Spectrosc.\/} {\bf 256} 57

\bibitem{docenko2010}
Docenko O, Tamanis M, Ferber R, Bergeman T, Kotochigova S, Stolyarov A~V,
  de~Faria~Nogueira A and Fellows C~E 2010 {\em Phys. Rev. A\/} {\bf 81} 042511

\bibitem{aymar2006a}
Aymar M and Dulieu O 2006 {\em J. Chem. Phys.\/} {\bf 125} 047101

\bibitem{guerout2010}
Gu\'erout R, Aymar M and Dulieu O 2010 {\em Phys. Rev. A\/} {\bf 82} 042508

\bibitem{kokoouline1999}
Kokoouline V, Dulieu O, Kosloff R and Masnou-Seeuws F 1999 {\em J. Chem.
  Phys.\/} {\bf 110} 9865--9877

\bibitem{kokoouline2000}
Kokoouline V, Dulieu O and Masnou-Seeuws F 2000 {\em Phys. Rev. A\/} {\bf 62}
  022504

\end{thebibliography}

\providecommand{\newblock}{}

\newpage

\newpage

\newpage

\newpage

\newpage

\end{document}